\documentclass[acmsmall,fleqn]{acmart}
\acmJournal{PACMPL}
\acmVolume{1}
\acmNumber{ICFP} % CONF = POPL or ICFP or OOPSLA
\acmArticle{42}
\acmYear{2020}
\acmMonth{1}
\acmDOI{} % \acmDOI{10.1145/nnnnnnn.nnnnnnn}
\startPage{1}

\setcopyright{none}
\makeatletter
\patchcmd{\@@addmarginpar}{\ifodd\c@@page}{\ifodd\c@@page\@@tempcnta\m@@ne}{}{}
\makeatother
\reversemarginpar

\usepackage{fancyvrb}
\usepackage{mathpartir}
\usepackage{listings}

\DefineVerbatimEnvironment%
  {core}{Verbatim}
  {xleftmargin=\mathindent}

\usepackage{cleveref}
\usepackage{subcaption}
\usepackage{caption}
\usepackage{tikz}
\usepackage{tikz-cd}
\usepackage{csvsimple,booktabs, siunitx, array}
\usepackage{sansmath}
\usepackage[utf8]{inputenc}
\usepackage{graphicx}
\usepackage{wrapfig}
\graphicspath{ {../img/} }
\begin{document}
\lstset{
  language=Haskell,
  basicstyle=\ttfamily\small,                % Code font, Examples: \footnotesize, \ttfamily
  numbers=left,                           % Line nums position
  numberstyle=\tiny,                      % Line-numbers fonts
  stepnumber=1,                           % Step between two line-numbers
  numbersep=5pt,                          % How far are line-numbers from code
  frame=none,                             % A frame around the code
  tabsize=2,                              % Default tab size
  captionpos=b,                           % Caption-position = bottom
  breaklines=true,                        % Automatic line breaking?
  breakatwhitespace=false,                % Automatic breaks only at whitespace?
  showspaces=false,                       % Dont make spaces visible
  showtabs=false,                         % Dont make tabls visible
  columns=fullflexible,
  keepspaces
}

% \begin{titlepage}

% This is the second revision of \textit{Fusing Industry and Academia at GitHub}. As
% per recommendation, it contains the following changes:

% \begin{itemize}

% \item The various future-looking sections have been corralled into individual paragraphs
% underneath an epilogue section (\ref{sec:conclusion}).

% \item The introduction of section \ref{sec:prodapps} has been reorganized and used
% as the seed of an additional subsection.

% \item Section \ref{sec:eff} has been rewritten and no longer uses a code sample, instead
% explaining the issue in prose.

% \end{itemize}

% \end{titlepage}

\title{Fusing Industry and Academia at GitHub (Experience Report)}

%% Author with single affiliation.
\author{Patrick Thomson}
% \authornote{with author1 note}          %% \authornote is optional;
                                         %% can be repeated if necessary
% \orcid{nnnn-nnnn-nnnn-nnnn}             %% \orcid is optional
\affiliation{
% \position{Position1}
% \department{Department of Computer Science}    %% \department is recommended
\institution{GitHub, Inc.}            %% \institution is required
% \streetaddress{Street1 Address1}
% \city{City1}
% \state{State1}
% \postcode{Post-Code1}
\country{United States}                    %% \country is recommended
}
  \email{patrickt@github.com}          %% \email is recommended

  \author{Rob Rix}
%   \authornote{with author1 note}          %% \authornote is optional;
                                         %%   can be repeated if necessary
%   \orcid{nnnn-nnnn-nnnn-nnnn}             %% \orcid is optional
  \affiliation{
%   \position{Position1}
%   \department{Department of Computer Science}    %% \department is recommended
  \institution{GitHub, Inc.}            %% \institution is required
%   \streetaddress{Street1 Address1}
%   \city{City1}
%   \state{State1}
%   \postcode{Post-Code1}
  \country{Canada}                    %% \country is recommended
}
  \email{robrix@github.com}          %% \email is recommended
\author{Nicolas Wu}
%\authornote{with author1 note}          %% \authornote is optional;
                                         %% can be repeated if necessary
%\orcid{nnnn-nnnn-nnnn-nnnn}             %% \orcid is optional
\affiliation{
%  \position{Position1}
% \department{Department of Computer Science}    %% \department is recommended
  \institution{Imperial College London}            %% \institution is required
%  \streetaddress{Street1 Address1}
%  \city{City1}
%  \state{State1}
%  \postcode{Post-Code1}
  \country{United Kingdom}                    %% \country is recommended
}
\email{n.wu@imperial.ac.uk}          %% \email is recommended

\author{Tom Schrijvers}
%\authornote{with author1 note}          %% \authornote is optional;
                                         %% can be repeated if necessary
%\orcid{nnnn-nnnn-nnnn-nnnn}             %% \orcid is optional
\affiliation{
%  \position{Position1}
%   \department{Department of Computer Science}    %% \department is recommended
  \institution{KU Leuven}            %% \institution is required
%  \streetaddress{Street1 Address1}
%  \city{City1}
%  \state{State1}
%  \postcode{Post-Code1}
  \country{Belgium}                    %% \country is recommended
}
\email{tom.schrijvers@kulueven.be}          %% \email is recommended

%

%% 2012 ACM Computing Classification System (CSS) concepts
%% Generate at 'http://dl.acm.org/ccs/ccs.cfm'.

%% End of generated code

\newcommand{\semantic}{\textsc{Semantic}}
\newcommand{\fe}{\textsc{Fused-Effects}}
\newcommand{\fastsum}{\textsc{Fastsum}}

%% Keywords
%% comma separated list
  %% \keywords are mandatory in final camera-ready submission

\begin{abstract}
  GitHub hosts hundreds of millions of code repositories written in hundreds
  of different programming languages. In addition to its hosting services,
  GitHub provides data and insights into code, such as vulnerability analysis
  and code navigation, with which users can improve and understand their software
  development process. GitHub has built \semantic{}, a program analysis tool
  capable of parsing and extracting detailed information from source code. The
  development of \semantic{} has relied extensively on the functional
  programming literature; this paper describes how connections to academic
  research inspired and informed the development of an industrial-scale program
  analysis toolkit.
\end{abstract}

%% \maketitle
%% Note: \maketitle command must come after title commands, author
%% commands, abstract environment, Computing Classification System
%% environment and commands, and keywords command.
\maketitle

\section{Introduction}

GitHub is a service that provides storage for repositories of source code
tracked with the Git distributed version control system. It is the largest
such service in the world, supporting over 65 million users and storing
petabytes of source code across hundreds of millions of repositories. The
size of the corpus of code on GitHub means that analyzing that code is a
source of significant business value. While GitHub boasts a large
engineering staff, we report the experience of Semantic Code, a team formed
in 2015 to create tools that analyze the corpus of open-source and
proprietary code stored on GitHub. One of these tools is a framework called
\semantic{}, a program analysis tool that supports diffing, code navigation,
and abstract interpretation. \semantic{} is implemented in the functional
programming language Haskell and is available as open-source
software.\footnote{https://github.com/github/semantic}

In order to extract up-to-date data from a user’s codebase, code analysis
services such as \semantic{} must operate whenever a user uploads a code
change to GitHub. This means that code analysis must be able to handle tens
of thousands of requests per minute, with thousands of simultaneous
connections, while producing useful data in a timely fashion. Such systems
present significant engineering and scaling problems. Industrial approaches
to the development of such systems are sometimes ad-hoc, but ad-hoc
approaches often suffer in terms of performance and comprehensibility, both
of which are hard requirements at GitHub’s scale. In order to avoid these
pitfalls, the Semantic Code team heavily draws on the literature associated with
functional programming research, including algebraic and scoped effects,
data types \`a la carte, recursion schemes, abstract definitional interpreters
and generalized LR parsing.

We have found that FP allows us to find
mistakes, test our assumptions, build prototypes, and to experiment within a
given problem domain. By leveraging techniques from this literature, the
Semantic Code team both solved pressing business problems and ended up with
production-tested libraries that we were able to release as open-source
software. This illustrates the bidirectional nature of exchange between
industry and academia: by drawing on academic techniques, industry can
contribute to software ecosystems at large.

In this paper, we describe the history of the \semantic{} project over the
seven years since the establishment of the Semantic Code team
and it initial prototype of the system (\Cref{sec:beginning}),
the techniques we used to scale this prototype so that
it could cope with production traffic (\Cref{sec:scaling}), and the
production applications powered by \semantic{} (\Cref{sec:prodapps}).
We then discuss the specific techniques for modelling effects that we have employed and refined
(\Cref{sec:evolving}), the varying levels of success
we have had with a range of functional programming techniques
(generalized LR parsing, algebraic effects, data types \`{a} la carte, recursion schemes)
and some of the lessons we have learned along the way
(\Cref{sec:lessons}).
Finally, we conclude (\Cref{sec:conclusion}).
We hope that this experience report
provides perspective on the scale of industrial problems, illustrates the
process of iterating on solutions within a given problem space, elucidates
real-world applications and connections to academic research, and affirms
that the academic community’s work is worthwhile and relevant to the
challenges faced in industrial software development.

% algebraic effects (\Cref{sec:effects}), recursion schemes
% (\Cref{sec:recschemes}), and data types à la carte (\Cref{sec:sharing},
% \Cref{sec:dtalc}), and how these advanced techniques fared in a real-world
% context under constraints of performance, time, maintainability and
% manpower. These techniques met with different levels of success, but even
% those that proved ultimately unsuitable provided us with fresh insight into
% the problem domain (\Cref{sec:codegen}).

\section{The Beginning} \label{sec:beginning}

\semantic{} has its roots in the realm of \emph{diffing}, a process of
determining a representation of a change to some source code. A \emph{diff} is a
representation, often textual, of a set of changes within a given software
project. The process of computing, applying, and displaying diffs is a
foundational responsibility of Git and other version control systems. However,
diffs of the sort emitted by the \texttt{git-diff} program are not always the
most readable rendition of a given change. Many diffs can span more than tens of
thousands of lines, with large diffs sometimes numbering in the hundreds of
thousands; as they increase in length, they tend to decrease in readability. The
\semantic{} project began in 2015, with a prototype of a new diffing algorithm,
intended to produce a more readable and informative diff than the basic
\texttt{git-diff} program---a \emph{semantic diff} that is aware of the
structural and syntactic qualities of programming languages. Such a diff would
use information derived from syntax trees to recognise structural changes in a
program, such as moving a function definition from one file to another, rather
than registering these changes as purely textual. (Note that a semantic diff
does not involve the denotational or operational semantics of a program.) The
fact that diffing is a core capability of version control systems, and of the
GitHub web application itself, motivated us to explore whether semantic diffing
could provide business value for users.

\subsection{Parsing}

Though GitHub hosts code written in thousands of different programming
languages, manpower constraints led the Semantic Code team to target a subset of
most popular programming languages on GitHub: Python, Ruby, JavaScript,
TypeScript, PHP, and Go. In order for \semantic{} to operate on this diverse set
of programming languages, we required a comprehensive approach to parsing and
analyzing source code. Real-world programming languages use varied techniques
and algorithms for parsing source code: the Ruby programming language uses a
LALR(1) parser generated with GNU Bison, whereas the Python language generates
its own parser out of a parsing expression grammar (PEG). The choice of parsing
algorithm becomes critical when considering syntactic structures that require
capabilities beyond that of some parsing algorithms. An example of this is
Python’s \texttt{with} statement, which requires multiple tokens of lookahead to
distinguish parenthesized expressions from multi-line grouped expressions in its
argument. CPython originally used an LL(1) parser, supporting only one token of
lookahead, which led to deficiencies in the
implementation\footnote{https://bugs.python.org/issue12782} that were only
remedied when the parser was rewritten in PEG style. We needed a parsing toolkit
that provided a consistent approach and application programming interface (API)
across several languages and also provided sufficient expressive power to parse
these languages correctly, regardless of the capabilities of their canonical
parsers.

Our chosen approach was built around the Tree-sitter parser generator
\cite{Brunsfeld18Tree}. Tree-sitter, a toolkit originally developed at GitHub to
provide syntax highlighting for the Atom text editor, uses the generalized LR
algorithm (GLR), first described by \citet{Lang74Deterministic} and first
implemented in 1984 by \citet{Tomita84Efficient}. The GLR algorithm can
recognise any context-free grammar, including ambiguous grammars and those
requiring arbitrary token lookahead. These capabilities allow Tree-sitter and
its grammars to serve as a \emph{lingua franca} for the world of programming
languages: regardless of the language under discussion, Tree-sitter is powerful
enough to recognise it, and its API is consistent across languages. Programmers
use a JavaScript domain-specific language (DSL) to express Tree-sitter grammars,
which generates a dependency-free C program, compilable to machine code or
WebAssembly, that any editor or programming tool can use to yield a syntax tree
from program text. By virtue of choosing Tree-sitter to power \semantic{}’s
parsing support, we were confident we could extend chosen approaches to any
language, as long as that language has a Tree-sitter grammar. The success of
Tree-sitter parsers in other applications and problem domains, such as GitHub's
syntax highlighting service, the Neovim text editor, and the Radare reverse
engineering toolkit, made us confident that the parsers themselves could
correctly handle languages as syntactically complex as TypeScript and Ruby.
Additionally, Tree-sitter parsers are tolerant of syntax errors: should a source
file contain invalid syntax, the error condition will be confined only to that
point in the syntax tree, and the remaining syntactically-valid code will be
present and accessible. This is a hard requirement, given that we cannot
assume all code in a repository is well-formed.

\subsection{An Initial Prototype}

Satisfied with our solution to the difficulties of cross-language parsing,
we wrote our initial prototype of a \texttt{semantic} executable in a beta
version of the Swift programming language\footnote{Accessible at commit
  \texttt{d23d646} of the \semantic{} repository.}. The Semantic Code team
had prior experience with Objective-C, Swift’s predecessor, which made it an
attractive platform given its additional type safety atop familiar
Objective-C APIs. While the prototype worked, it suffered from poor
performance and poor developer experience. Though the diffing algorithm was
clearly not optimized yet, there was too much friction in writing
deployment-ready code in Swift: given its beta status Swift was evolving
rapidly and its rapid language changes and sometimes-unstable toolchains
distracted from larger engineering goals. Since our code was written in a
functional style, we looked for more established languages that would
preserve the functional style without compromising performance or
readability. Haskell seemed an appropriate choice, given its decades of
history, its mature, native-code Glasgow Haskell Compiler (GHC), and its
success in other industrial settings such as Meta’s Sigma anti-spam system
\cite{marlow15spam,marlow14fork}. Moreover, Haskell’s foreign-function
interface (FFI) made it easy to link to Tree-sitter generated parsers.

We were astonished by the speed with which we converted our Swift code to
Haskell\footnote{Accessible at commit \texttt{95c2850}.}. After less
than a day’s engineering effort, our Haskell implementation outperformed the
Swift code by a factor of two; this was doubly remarkable given that this was
the team’s first experience writing Haskell in industry. GHC’s FFI support
allowed us to operate on syntax trees via the C-based Tree-sitter API, and we
generated Haskell data types suitable for diffing and analysis by passing these
syntax trees as the seed value to an anamorphism \cite{Meijer91Bananas}.

\section{Scaling the Prototype with Functional Techniques}
\label{sec:scaling}

Buoyed by the ease of implementing this prototype with Tree-sitter and with
Haskell, we turned to the next issue facing us: how were we to take this
prototype and build something capable of handling GitHub’s stringent engineering
requirements and considerable production traffic? Scaling a prototype is not
just a matter of performance: it also involves planning to keep code complexity
under control. A single programming language is complicated enough to implement
and analyze; we feared that the complexity associated with a naïve approach to
implementing a multi-language analyzer would impair the development of any
interesting analysis features.

\subsection{Syntactic Sharing} \label{sec:sharing}

Both empirical studies \cite{Haefliger08Code} and the team’s anecdotal
experience indicate that code reuse is an effective method for managing
complexity. The team decided to achieve a degree of code reuse by sharing the
representation of ASTs across our target programming languages. As an example of
this, many languages share syntactic features that are relatively similar, such
as simple arithmetic operations, functions, assignment statements, and comments.
Manually defining a \texttt{Comment} syntax type for each language would be both
tedious and complicated. Other language variants overlap more substantially,
such as in the case of TypeScript, which is a superset of JavaScript: given
their degree of shared semantics, it was our goal to reuse analyses written for
TypeScript on JavaScript codebases.

Our chosen solution for syntactic sharing would let us compose a language’s
syntax types out of smaller, shared parts. We decided to use a data types à la
carte \cite{Swierstra08Data} methodology, which meant defining ASTs as a
coproduct of endofunctors, each representing the shape of a kind of node (a
string literal, a function call, etc.), tied into a recursive shape with a
simple fixpoint. Another fixpoint, defining recursive positions as either a
copy, insertion, deletion, or replacement of ASTs, gave us a representation for
diffs. All syntax types were functors parameterized in terms of an additional
data type representing a node's annotation; this polymorphism allowed us to
track source locations and to annotate, in the case of diffing, whether a node
was added or removed. Syntax errors in the tree were represented with a shared
error type.

\begin{lstlisting}[label={code:dtalc}, keywords={type}, title=A representation of JSON syntax written with data types à la carte.]
  import Data.Sum (Sum)
  import Control.Comonad.Cofree (Cofree)
  import qualified Syntax
  import qualified Syntax.Literal as Literal

  type JSONSyntax =
  [ Literal.Null,
    Literal.List,
    Literal.Boolean,
    Literal.Hash,
    Literal.Decimal,
    Literal.KeyValue,
    Literal.TextElement,
    Syntax.Error -- for ill-formed nodes
  ]

  type JSONTerm = Cofree (Sum JSONSyntax)
\end{lstlisting}

An advantage of representing syntax trees as fixed points of coproducts of
endofunctors is that such a representation is compatible with recursion schemes
\cite{Meijer91Bananas}. The team employed recursion schemes whenever the shape
of data permitted it, as in our experience operations written with recursion schemes
are often more flexible, readable, and type-safe than those written with
explicit recursion. We used catamorphisms and anamorphisms to implement
operations that generated, manipulated, and serialized syntax trees, and used
paramorphisms for operations such as term rewriting and dead code elimination
that required additional context.

Every engineering decision comes with trade-offs, and choosing data types à
la carte provided us with compositionality and fluent recursion but
sacrificed a degree of type safety: because these types are functorial with
respect to their children, we cannot constrain the types of the children
without giving up functorial map. Another downside of this representation
was that it required a preprocessing stage, dubbed \emph{assignment}, an
unparser that converted a Tree-sitter syntax tree to a sum-of-products
Haskell type, based on an anamorphism implemented once per desired language.
This code resembled a construct dual to the Tree-sitter specification of the
parser, but had to be written manually. Despite these drawbacks, we grew
comfortable with the use of coproducts of functors to represent ASTs,
turning then to improving the performance of the program itself.

\subsection{Algorithmic Improvement}

Our algorithm was initially simple, essentially treating a given syntax node as
a tuple, a list, or a dictionary. We diffed tuples—nodes with a fixed set of
children—in $O(n)$ time by diffing corresponding members. Lists represented
nodes with arbitrarily many children, and were diffed with an approach based on
the classic shortest edit script (SES) problem \cite{Myers86Diff}, initially
using a naïve, compare-everything-to-everything-else algorithm running in
$O(n^2)$ time. Dictionaries, mapping keys to values, were diffed via set
reconciliation of the keys. All of this was wrapped up in a small DSL
implemented using a free monad, giving us a high-level vocabulary capable of
understanding and executing diff scripts generated automatically or by hand,
supporting all of our target languages. Breaking syntax down into sums of small
syntax functors eventually allowed us to define what we called
\emph{sub-structural diffing}, where a piece of syntax can nominate some
sub-syntax as mediating its identity; this allowed us to diff functions by
comparing their identifiers.

In 2016, performance concerns and a desire to eventually detect code moves and
renames, and other non-minimal features of readable diffs, led us to borrow
parts of the RWS-Diff algorithm \cite{finis13rwsdiff}, which applies techniques
from computer vision to the problem of comparing trees. While we found this
suboptimal due mainly to its use of a pseudorandom number generator and its
consequent unpredictability, it did improve our algorithm's efficiency,
particularly when combined with other approaches. For example, we decided to
improve the naive $O(n^2)$ diffing algorithm that we had written originally for
variable-arity branches, and eventually as a preflight pass applied before RWS.
The diffing system was improved by implementing an $O(nd)$ time algorithm by
\citet{Myers86Diff} where $d$ is the size of the difference, which was
extremely efficient, particularly in the expected case that more has remained
the same than has changed. Fixed-arity branches continued using sequential
diffing, running in linear time.

\section{Production Applications} \label{sec:prodapps}

Having achieved acceptable performance, our next course of action was to
replace the diffs shown on the \href{https://github.com}{github.com} website,
particularly on pull request views, with syntactic diffs emitted from
\semantic{}. This section explains why we had to abandon this approach for
non-technical reasons, and how our goal shifted instead to applications
related to source code navigation and comprehension.

\subsection{Production Showstoppers for \semantic's Diffs}
We had come to a point where \semantic{} was able to output its diffs in a format
compatible with \texttt{git}. We could therefore integrate these diffs into the
\href{https://github.com}{github.com} interface as a drop-in replacement for
files in supported languages. However, non-technical factors led us to abandon
this course of action: specifically, the fact that diffs on GitHub would differ
from those generated by the standard \texttt{git-diff} program was deemed an
insurmountable barrier to widespread adoption. Although Git itself can be
configured to use syntactic diffs emitted from a \texttt{semantic} binary, it
would require users to download, install, and understand an extra tool, one
less battle-tested than \texttt{git-diff} itself.

We then turned to another feature. With diffs represented as syntax trees where
some nodes have been replaced by patches, we had all the information we needed
to compute summaries of the patches occurring within a diff in a high-level,
readable description of what had changed. Unfortunately, while collecting the
changes was a matter of a trivial fold, producing a high-level, intelligible
description from these proved to be much harder, especially given that these
were changes to source code which itself is difficult to summarize. It turns out
to be quite a challenge to do better than just presenting the changes verbatim,
at least for a user base largely consisting of experienced programmers. We arrived
at no formulation that substantially improved on the act of reading a textual diff.

\subsection{Table-of-Contents Analysis}

\begin{wrapfigure}{r}{0.5\textwidth}
  \begin{center}
    \includegraphics[width=0.4\textwidth]{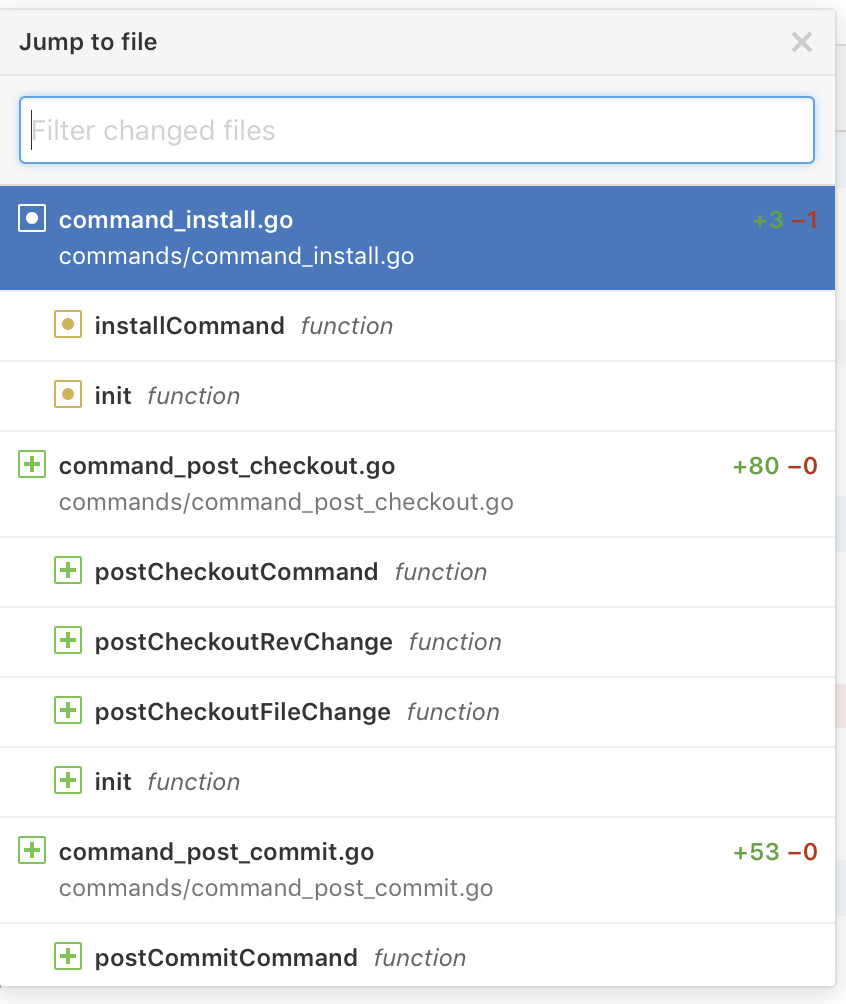}
  \end{center}
  \caption{The GitHub table of contents feature as rendered to users.}
  \label{fig:TOCScreenshot}
\end{wrapfigure}

Our subsequent engineering efforts focused on a \textit{table of contents}
feature. The table of contents associated with a given diff provides a summary
of the files and functions changed in that diff. This feature was shipped to the
public \cite{rix17changed}. A remote-procedure call (RPC) server called
\texttt{semiotic}, written in Go, listened for requests for a table of contents,
executed the \texttt{semantic} binary out-of-band to analyze that diff, and
returned the table of contents data over the wire, encoded in Google Protocol
Buffers format \cite{google08protocol}. This information was then decoded by the
Ruby on Rails application that powers \textsf{github.com} and rendered
appropriately in its interface (Figure ~\ref{fig:TOCScreenshot}).
Though this feature performed well and provided users with useful information,
it did not see wide user adoption, possibly due to its limited space in the pull
request UI. GitHub now has a side-bar view, inspired by the table of contents
feature, that is displayed on all pull requests.

The fact that this service yielded useful customer data at GitHub levels of
scale considerably increased the project's momentum. One of the Semantic Code
team's central goals is to build solutions that require zero configuration from
users: GitHub already provides a powerful analysis tool in CodeQL, and supports
custom analysis tools in continuous-integration pipelines implemented with
GitHub Actions, but these solutions require manual configuration for each
repository in order to integrate with users' build processes. At this point, our
goal shifted from diffs and diff analysis to the navigation and comprehension of
source code itself, without requiring the repository owner to perform any work
to yield its benefits.

\subsection{Code Navigation}

Our first effort after the table of contents feature was to improve an internal
experimental prototype code navigation system. The archetypal code navigation
feature, common to many text editors and IDEs, is \emph{jump to definition}.
This allows a user, when faced with an unfamiliar function or variable, to
travel immediately to its definition. Dual to this feature is \emph{find all
  references}, which, given the definition of a function or variable, locates
all parts of the program that reference or invoke that entity. Given that GitHub
users often use the \textsf{github.com} web interface to browse code, we
anticipated that code navigation as a GitHub feature would be profoundly helpful
to those trying to navigate and understand large or unfamiliar codebases.

The prototype code navigation system was used internally by GitHub staff, and
proved a helpful tool in the maintenance of the large and complicated Ruby on
Rails codebase that underlies GitHub itself. While this prototype worked well
enough for internal purposes, there were concerns that it would not scale to a
production system visible to all GitHub users. The original system was written
in Ruby and extracted data from source code by invoking and parsing the output
of the external \texttt{ctags(1)} Unix utility. We wanted to bring reliability,
performance, and breadth of utility to this service, without the user having to
configure any aspect of the system, and decided to build a prototype of this
feature atop our Haskell codebase. We considered an approach built on the code
navigation capabilities of external language tooling, over the Language Server
Protocol (LSP) standard, but discarded it due to the operational difficulty
associated with deploying a separate software stack for each targeted language,
as well as the impedance mismatch between LSP capabilities and our tree-oriented
view of the world.

\begin{wrapfigure}{r}{0.5\textwidth}
  \begin{center}
    \includegraphics[width=0.48\textwidth]{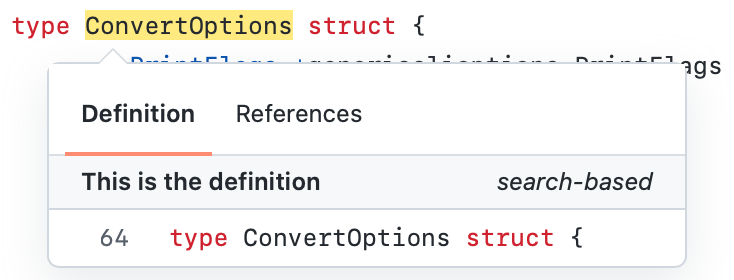}
  \end{center}
  \caption{The GitHub jump to definition user interface.}
  \label{fig:CodeNavScreenshot}
\end{wrapfigure}

Though sharing the syntax tree provided a high degree of reusability when
implementing a \texttt{ctags} analysis for several different programming
languages, the complexity of these languages reared its head again. An example
of this is Ruby’s support for allowing function calls that omit parentheses in
parameter lists, similar syntactically to invocation of Unix shell commands.
This syntactic quirk means that it is not possible to distinguish syntactically
between a zero-argument function call and a reference to a variable. Analyses
must keep tables of local variable declarations and use these tables to
disambiguate them from zero-argument function calls.

Implementing code navigation was a matter of defining a type class that emitted
tag information and implementing this class for every relevant node. This was
made available through \texttt{semanticd}, an HTTP-based RPC service, written
with the Servant web API library \cite{servant14} and linking in \semantic{} as
a library, that wrote tag information to a MySQL database upon pushes to GitHub
repositories and fetched that information based on user queries. The resulting
system performed admirably when deployed to production, handling tens of
thousands of requests per minute, deployed and scaled via the Kubernetes
\cite{google14kube} container management system. \texttt{semanticd} proved
reliable at GitHub scale: the primary source of crashes was a vendor-specific
hardware bug that the GHC maintainers promptly
remedied.\footnote{https://gitlab.haskell.org/ghc/ghc/-/issues/18033}

\subsection{Semantic Analysis}

With the infrastructure for simple code analysis such as diffing and table of
contents in place, the next goal of the project was to perform more
sophisticated code analysis on repositories to drive features for end users. The
goal was to have multiple forms of analysis over multiple different languages.
Without a principled approach to modularity, this problem would easily become an
engineering nightmare: if supporting $m$ features for $n$ languages requires $m
\times n$ amount of work, it becomes increasingly untenable to support new
features on languages with a small team, and completely unreasonable to expect
the same from third-party contributors. Focusing our attention on the top five
most-used languages is already a problem, as general-purpose languages are often
quite large, and language size is an important factor in the effort required.

An optimal solution would reduce this to an $m + n$ problem where each new feature
or programming language need only be implemented once. Now the effort required
by the team is much more tractable, and third-party contributors can more
reasonably add support for new languages.

Further, since both our team and teams likely to be clients of these services
lacked prior experience with program analysis, our solution needed to be approachable;
we expected to serve a variety of different end-user–facing products, and so we
needed it to scale to multiple analyses; and we knew we would need to be able to
experiment to tune both performance and precision of the results to suit the
constraints of each use case.

Finally, while we could already serve some needs by means of syntactic analyses,
typically implemented as folds over syntax trees and diffs, properties which
depended on any of a program's dynamics, for example computing the set of
exception types that can be thrown through a given call site, or collecting the
set of instructions unreachable from a given set of inputs, remained outside of our
reach. Therefore, we knew we needed a truly semantic analysis, justifying for the
first time our toolkit and team's heretofore aspirational names.

Serendipitously, it was at about this time that the team recognised that work by
\citet{Darais17Abstracting} on Abstracting Definitional Interpreters (ADI) was
relevant to these problems. All ADI requires of us is an evaluator written in an
open-recursive style against a set of capabilities based on the work of
\citet{Horn10Abstracting}. Implementing an evaluator for an entire language is a
non-trivial task, but allows us to support the vast majority of analyses, and
hence end-user features, without additional effort, reducing our $m \times n$
problem to $m + n$ at a stroke. Furthermore, motivated third parties such as
communities of language users can contribute to an evaluator, just like they
already do with parsers. This reduces the burden on our team, while often also
allowing bugfixes to be provided with a faster turnaround than could
otherwise be the case.

Further, ADI gives us a large number of levers to control the performance and
precision of analyses. Conveniently, these are the two metrics by which we tune
an analysis. For example, the monadic nature of the evaluator's targeted
interface allows semantics to be altered simply by switching the ordering of the
handlers of the various components of the abstract machine. (Note that in the
paper, these are implemented via monad transformers, while our implementation
employs algebraic and scoped effects \cite{wu2014effect}.)

To some degree, this began as a solution in search of a (specific) problem: it
was clear that we would need this suite of capabilities for future developer
productivity features, but it wasn't at all obvious what they should be. At one
time, we planned on implementing code navigation using abstract interpretation,
but in the end that implementation was pursued in a Rust service using a notion
of \emph{stack graphs} \cite{creager21stackgraphs} inspired by scope graphs
\cite{visser18scopes}, which allowed us to express scope-aware code navigation
without the time and effort associated with developing full abstract
interpretation for all targeted languages. This finer-grained solution proved
more expedient for us given our time constraints, and by implementing it as an
addition to tree-sitter allowed us to avoid the overhead of serializing between
Tree-sitter and \semantic{}. However, we continue to explore definitional
interpreters for analyses more sophisticated than those required for code navigation.

\section{Evolving Approaches}
\label{sec:evolving}

\semantic's successful production deployment confirmed our belief that
academic research was an effective source of techniques and approaches. However,
working in a problem space as complex and diverse as analysis of real-world
programming languages presented confounding factors to which we had to adjust.
Some of these factors stemmed from runtime performance issues, some from
compile-time performance, and some from our desire to manage complexity more
fluently.

\subsection{Effective Haskell} \label{sec:effects}

As our grasp of syntax became more precise, our need to express the
capabilities of operations on these types grew: we wanted to delineate what
operations could and could not happen where. We found that the complexity
associated with rigorously expressing capabilities with the Haskell type system
was justified, given the clarity that these types provided and the errors that
they prevented. We found that one of the appeals of Haskell in an industrial
setting is that it is clear in showing what you can and cannot do.

\newcommand{\mtl}{\textsc{Mtl}}

Haskell programmers most commonly use a finally-tagless approach
\cite{Carette07finallytagless} when expressing
the capabilities of code in the type system using the monad transformer library
\mtl{}, to express computational effects \cite{Jones95Functional}. This approach
suffers in that it requires a separate monad transformer for each effect’s
interpretation, and each monad transformer must implement an instance per
effect, which became infeasibly complex given that our most complicated analyses
sometimes involved invocations dozens of effects deep. We reached instead for
algebraic effects \cite{Plotkin01Semantics}, a family of techniques that,
instead of using type class instances in the manner of \mtl{}, represent effects
with data constructors and handlers that interpret invocations of these
constructors. We forked and modified an existing effect system developed by
\citet{freersimple16}. Our successful experience with the Kiselyov and Ishii
formulation of algebraic effects established it as a foundational tool for
subsequent work in research and in production systems.

We saw an early example of the utility of algebraic effects during the
development of our diffing algorithm. During the debugging process, we needed to
dump the state of the $O(nd)$ algorithm during its operation. In order to do
this, we implemented effect handlers that both performed the diffing algorithm
and emitted its intermediate states as SVG (Scalable Vector Graphics) files.
Effect handlers made it straightforward to implement the algorithm and allowed
us to customize its interpretation when we needed it to emit additional data.

\subsection{Effects and Interpretations} \label{sec:eff}

Several issues led us to conclude that we needed an effect system which could
represent not only algebraic effects, but also effects which themselves contain
effectful actions.

% \paragraph{Surprising results when using effects from within handlers.}
% If a handler uses
% an action that would be modified by a surrounding operation's scope, the change in
% question will not be apparent. For example, the \texttt{Reader} effect models a
% variable using the operations \texttt{ask}, which returns the value to which the
% variable is bound, and \texttt{local}, which shadows the variable's binding within
% its scope.
%
% \begin{wrapfigure}{r}{.6\textwidth}
% \vspace{-2ex}
% \begin{lstlisting}[keywords={}, numbers=none]
% runReader r (Ask k) = runReader r (k r)
% local f (Ask k) = ask >>= local f . k . f
% runPing (Ping k) = ask >>= print >> runPing k
%
% > runReader 0 (runPing (ping >> local (1 +) ping))
% \end{lstlisting}
% \vspace{-12pt}
% \caption{\texttt{Reader} effect pseudo-code.}
% \label{code:reader}
% \end{wrapfigure}
%
% In Figure~\ref{code:reader}, a \texttt{Ping} operation, which returns to the handler
% without expecting any value in response, is given a handler which prints a
% variable's value whenever a ping is received. The action sends a ping, and then
% another ping inside a locally- incremented scope, with the intention of
% outputting \texttt{01}. However, the output is surprising: it prints \texttt{00}
% instead, as though the call to local had never occurred.
%
% While we want the body of the \texttt{Ping} handler to be substituted for the ping
% operations, it is instead its \emph{result} which is substituted. Thus,
% \texttt{local} never sees the \texttt{ask} inside it, and thus cannot increment the
% value printed.

Some effectful actions take effectful computations as parameters. One example is the
\texttt{Reader} effect, which is equipped with two primitive actions:
\texttt{ask} returns the ``environment'' value and \texttt{local $f$ $p$}
modifies the environment value, using the function $f$, but this modification
is local to the computation $p$. The standard algebraic effects approach implements
\texttt{ask} as a constructor and \texttt{local} as an interpreter.

As pointed out by~\citet{wu2014effect}, when combined with other effects,
implementing \texttt{local} and similar operations as interpreters leads to
surprising, and often undesired behavior. We ran into this hard-to-debug pitfall
on multiple occasions.
%
% \paragraph{Difficulty embedding actions in effect types.} One way to understand the above
% issue is that \texttt{local} is an \emph{interpreter} for \texttt{ask} operations,
% not an effectful operation itself. However, resolving this is not as easy as it
% might sound. \texttt{local} belongs to the effect \texttt{Reader r}, but it contains
% an action which must itself be able to perform \texttt{Reader r} effects, as well as
% any other effects which can be performed where \texttt{local} is called, so it can
% pass these through to other handlers. The signature of effects which can be performed
% is communicated through types, and this in turn means that the type of this action
% must therefore contain the type to which \texttt{local} belongs—a cyclic definition
% which is difficult if not impossible to write down, let alone use.
%
% \texttt{local}, and other effects containing effectful actions,
This
convinced us of the need for
an effect system capable of modelling not just algebraic effects, but also
non-algebraic \emph{scoped} effects—effects which delimit the scope within which
some behaviour---locally shadowing a variable, catching an exception, forking a
computation using cooperative multitasking, etc.---will be applied.

Though \mtl{} is the \emph{de facto} effect system for developing Haskell
applications, it was unsuitable for implementing the system described by
\citet{Darais17Abstracting}, as the implementation relies on the ability to have
multiple state types, with multiple interpretations, in a given monadic
computation. \mtl{} prohibits this approach due to the fact that its effect type
classes, to aid in type inference, prohibit multiple instances of these classes
for a given monad transformer. The recommended solution for this is to create
more monad classes, and to implement these classes for all monad transformers,
which is the $O(n^2)$-instances problem discussed previously. Despite this
drawback, \mtl{} performs very well, orders of magnitude better than a
freer-monad approach, due to the fact that the GHC optimizer is eager to inline
type class methods \cite{PeytonJones02Secrets}, avoiding the overhead of
constructing type class dictionaries. However, GHC is generally reluctant to
inline recursive code, and the act of handling algebraic effects is inherently
recursive, given that one handler might need to delegate to another. Our
solution needed to introduce as few performance changes as possible, both out of
performance requirements and ability to implement effects such as
telemetry-based performance monitoring: adding effects needed to avoid altering
the performance characteristics of the code under examination. As such,
we had to recover a degree of the inlining characteristics of \mtl{}.

\subsection{\fe{}}

The aforementioned deficiencies in our freer-monad effect system caused us
to look at alternatives. We looked, again, to the literature to help us
overcome these problem. Work by \citet{wu2014effect} provided a formulation
of effect handlers capable of handling scoped operators. \citet{wu15fusion}
demonstrated that GHC's reluctance to inline recursive effect handlers is
remedied by expressing these handlers as type class methods. Additionally,
work by \citet{Schrijvers19Monad} allowed us, by limiting ourselves to
modular effects, to remove occurrences of the freer monad entirely, instead
using the well-attested and GHC-friendly monad transformer approach. We
implemented \fe{}, an effect system based on these techniques, and
immediately observed considerable speedups in analysis benchmarks. The \fe{}
support for scoped effects with multiple interpretations rendered it trivial
to implement effects such as telemetry. We have released \fe{} as
open-source
software\footnote{https://github.com/fused-effects/fused-effects}, as well
as associated packages providing integration with common libraries such as
\texttt{haskeline} and advanced approaches to data manipulation such as
profunctor optics \cite{Pickering17Profunctor}. With more than 10,000
downloads to date, \fe{} has seen substantial community adoption and
industrial use outside GitHub.

\section{Lessons Learned}
\label{sec:lessons}

We consider functional programming to have been a clear win for \semantic{}.
In practice, some techniques applied better than others; nevertheless, we learned
a great deal in the process of applying and evaluating a variety of approaches.

\subsection{GLR Parsing}

Tree-sitter and the generalized LR algorithm it provides have proven
rock-solid at industrial scale and sufficiently expressive to parse even
languages, such as Ruby and TypeScript, that display tremendous syntactic
complexity. Tree-sitter's JavaScript-based DSL is sufficiently expressive to
cover most aspects of PL syntax, and for languages with complex lexing
rules, such as Ruby's support for nested heredoc strings, Tree-sitter allows
hand-written lexers in C or C++. On the occasions we had to extend
Tree-sitter's capabilities, such as to emit information about a grammar's
syntax types for code generation (see Section \ref{sec:codegen}), doing so
proved straightforward. The Tree-sitter ecosystem continues to expand,
thanks to its significant community adoption, and we anticipate that it will
continue to serve as a fundamental building block for the Semantic Code
team's engineering efforts. The fact that language communities themselves
can maintain their language's Tree-sitter grammar allows GitHub to provide
useful code intelligence features, without having to commit GitHub
engineering time to language support itself. Given the plethora of
programming languages hosted on GitHub, this is an elegant solution to
manpower and time constraints, one that also benefits language communities.

\subsection{Algebraic Effects}

From the project's inception, our use of algebraic effects was pervasive and
profound. We conjecture that \semantic{} is among the largest Haskell projects
developed without any direct dependency on \mtl{}. We found algebraic effects an
elegant, effective, and expressive way of writing effectful code in Haskell.

The monadic computations described in \citet{Darais17Abstracting} require
multiple \texttt{State} and \texttt{Reader} types in their transformer stacks,
and the paper's artifacts accomplish this by defining a bespoke monad
transformer library, alongside macros to ease the composition of monad stacks.
Expressing these computations using \mtl{} would be tedious, as every duplicated
\texttt{State} or \texttt{Reader} effect would need to be wrapped in its own
finally-tagless monadic interface, alongside a concrete transformer type that
would require instances for all associated monad transformer classes. This is
the classic $O(n^2)$ instances problem at work. Because \fe{}'s effect
invocations are more polymorphic than those provided by the \mtl{} interface,
this posed no difficulty, allowing us to port the aforementioned interpreters
from the Racket implementation in \citet{Darais17Abstracting} to Haskell with a
minimum of fuss and effort. Though \mtl{} yields better type inference thanks
to the functional dependencies of its effect classes, GHC's support
for visible type applications \cite{eisenberg16visible} allowed us a pleasing
syntax for disambiguating any ambiguous invocations.

An example of the utility and flexibility of algebraic effects is an effect we
developed to extract telemetry data from our production Haskell systems. A
hard requirement for production systems is that they emit data about the state
of the system, in order for their maintainers to have insight into the
behavior of a program during the operating and debugging process. Examples of
such include log messages and associated key-value data; remote metrics such
as counters, timers, and statistical distributions; and execution traces that
describe the control flow paths taken by a given request or action. These data
are sent to internal services that store, aggregate, categorize, and display
them in a manner that aids comprehension. Algebraic effects provided us a rich
vocabulary for aggregating and collecting these data, as well as the ability
for this aggregation and collection to operate differently in different
contexts.

Timing the execution of a given code block is an example of a scoped effect: the
yielded telemetry data are limited to the code around which the timing invocation
is wrapped. However, sometimes we wanted different behavior, especially during
local development: data aggregators are intended for production, rather than
development, but the statistics generated are still useful and relevant during
the development cycle. In a production context, we wanted these data to be
uploaded to an aggregator; in a development context we wanted to see them
reported on the command-line; and when running automated tests, we wanted to
discard them entirely. Furthermore, our solution needed to introduce as few
performance changes as possible, lest the act of measuring some code's execution
time alter the performance characteristics of the code under examination. With
\fe{}, it was trivial to define a \texttt{Telemetry} effect and associated
interpreters to handle the three above cases, thanks to \fe{}'s support for
reinterpreting scoped effects. A simpler effect system that did not support
such reinterpretation would not have sufficed.

\subsection{Data Types à la Carte} \label{sec:dtalc}

Our initial experience with data types à la carte was positive. Expressing a
language's syntax with a type-level list (see Section \ref{code:dtalc}),
thanks to GHC's support for type-level programming, and then parameterizing
the associated type with that list, provided an elegant and uniform
interface to querying and analyzing syntax trees. The fact that we shared
common syntax nodes, like \texttt{Comment} and \texttt{Integer} types,
between languages allowed us to generalize functions and analyses across
said languages. Performance problems associated with a recursively defined
notion of subsumption\footnote{https://gitlab.haskell.org/ghc/ghc/-/issues/8095} led to us
developing a library called \fastsum{}, released as open-source
software\footnote{https://github.com/patrickt/fastsum},
that unrolled this recursive loop within the Haskell type system, which bypassed
performance problems associated with large languages such as TypeScript, which
contains hundreds of distinct syntax nodes.

The utility of the data types à la carte approach was ultimately limited due to
the variance in semantics between languages with what superficially appeared to
be the same kinds of syntax. For example, object-oriented programming languages
that support implementation inheritance provide a construct that instructs the
language to invoke a superclass method. In Python, Java, and Ruby, this
construct is written \texttt{super()}, suggesting that they should be modeled by
a single syntax type. However, Ruby provides an unusual variant of
\texttt{super()}, known as “zsuper”, short for “zero-argument super” and written
without trailing parentheses. When invoked in a method taking one or more
arguments, zsuper implicitly forwards all arguments in scope to its superclass’s
implementation. We defined a Ruby-specific zsuper syntax functor, and made
similar compromises when dealing with other quirks of real-world programming
languages: ultimately, this problem domain is sufficiently complicated to, at
times, actively resist abstraction.

Additionally, an à la carte approach is ultimately unityped: even though syntax
trees generally permit type invariants on their subtrees, our approach
discarded this information. Though this made it easy to represent nodes that
can appear anywhere, such as comments or parenthesized expressions, it
complicated tree rewrites, as we had to anticipate the possibility of such
occurrences at any given position in the syntax tree. For example, an operation
that requires portions of a syntax tree may be valid in a case without comments
present in the syntax tree, but if comments can appear at any point in a syntax
tree, that operation must account for the presence of a comment as well as
preserve its position in the syntax tree. Correctly accounting for edge cases
such as these involved many runtime type checks, which obviated many of the
gains from choosing a strongly typed implementation language. Furthermore, the
assignment stage, the language-specific anamorphism mapping a Tree-sitter tree
to a coproduct of functors, proved a significant maintenance burden: any update
in a language grammar required modifications to the assignment stage. Assignment
code had to be written manually, held us back from upgrading language grammars
regularly, was difficult to debug, and proved to be a reliable source of bugs.

\subsection{From à la Carte to Code Generation} \label{sec:codegen}

The team ultimately chose to move away from an à la carte representation of
syntax trees, electing instead to generate Haskell data types directly from type
information emitted by the Tree-sitter parser generator. This approach,
described and implemented by \citet{Nadeem20CodeGen}, proved superior in several
aspects: it retained the type information that the à la carte representation
lacked, the generated types required no maintenance by hand, and it eliminated
the need to implement an assignment stage altogether! Because syntax trees
yielded from Tree-sitter may not be well-formed, the fields of syntax types are
wrapped in an \texttt{Err} functor isomorphic to \texttt{Maybe}.

\begin{lstlisting}[label={code:codegen}, keywords={type,data,newtype},
  title={
    A simplified representation of the syntax types generated
    from the Tree-sitter grammar for JSON.
  }
  ]
  -- Terminal nodes and their textual contents
  data True a = True {ann :: a, text :: Text}
  data False a = True {ann :: a, text :: Text}
  data Number a = Number {ann :: a, text :: Text}
  data Null a = Number {ann :: a, text :: Text}
  data String a = Number {ann :: a, text :: Text}

  -- Nodes with children
  data Array a = Array {ann :: a, children :: [Err (Value a)]}
  data Pair a = Pair {ann :: a,
                      key :: Err ((Number :+: String) a),
                      value :: Err (Value a)}
  data Object a = Object { ann :: a, children :: [Err (Pair a)]}

  -- Toplevel choice node representing JSON values
  newtype Value a = Value
    ((True :+: False :+: Number :+: Null :+: String :+: Array :+: Object) a)
\end{lstlisting}

Ultimately, the utility of reusing grammar descriptions outweighed the utility
of reusing hand-maintained syntax types. Though a code-generation approach
creates more syntax data types than does an à la carte representation, the
elision of the complicated and error-prone assignment stage made this a sensible
tradeoff in light of our engineering requirements. Despite à la carte syntax
proving ultimately unsuitable, it showed that code reuse is more effective when
it integrates with high-level data descriptions, such as the JavaScript DSL
that specifies a grammar description.

\subsection{Recursion Schemes} \label{sec:recschemes}

Recursive operations are a fundamental building block of idiomatic Haskell, and
as such a toolkit providing a generalized vocabulary for many different types of
recursion proved helpful for a wide array of tasks. The Haskell \texttt{recursion-schemes}
library provides a sophisticated vocabulary for catamorphisms, anamorphisms,
paramorphisms, and histomorphisms, alongside an API that integrates well
both with standard Haskell data structures and with our hand-written syntax
types, which, being functorial with respect to their subterms, were already
compatible with the standard fixpoint encoding of codata in Haskell. An example
of recursion schemes' applicability came during the prototyping of a DSL for
term rewriting; expressing the DSL's interpreter with a paramorphism elided all
explicit recursion and generalized the DSL to any data type compatible with
\texttt{recursion-schemes}. Even the more exotic recursion schemes proved
themselves useful, such as histomorphisms during our experiments with diff
summaries and hylomorphisms as part of our implementation of RWS-Diff.

The problems with recursion schemes emerged most notably when transitioning to
the generated syntax types described in Section \ref{sec:codegen}. These generated types are
functorial in terms of their annotation type rather than their subterms, and as
such the traditional encoding of recursion schemes via a fixed point of functors
does not apply. There exist approaches that generalize the data types à la carte
to mutually-recursive, well-typed trees \cite{bahr11compositional}, but the
incidental complexity is high, and the sheer size of syntax trees like those of
TypeScript precluded traditional encodings of sums and subsumptions (hence our
development of \fastsum{}). Defining a higher-order equivalent of the
traditional \texttt{Traversable} type class, as well as the required code to
derive these instances generically \cite{Magalhaes10generic} recovered a
degree of expressivity when recursing into subterms, though at the cost of
flexibility when compared to the à la carte formulation.

\section{Epilogue}
\label{sec:conclusion}
This section summarizes \semantic's current situation and future plans,
and then concludes.

\paragraph{Current Situation}
While the Haskell implementation of code navigation was a success, it was
later replaced by a domain-specific language for querying syntax trees. This
decision did not stem from deficiencies in \semantic{} itself: rather, our
manpower constraints precluded the use of heavy-duty language stacks. Due to
the considerable number of programming langauges in use, we needed to enable
external contributors to maintain their own code navigation rules without
having to write Haskell code. Haskell’s considerable ecosystem, large
compiler, and learning curve made it an unsuitable choice for language
maintainers unfamiliar with Haskell. (The same is true of Rust, Tree-sitter's
implementation language). Our goal was to eliminate all possible
barriers to entry. Because the query language we developed is useful in
other contexts, we implemented it directly in Tree-sitter, adding to its
existing suite of tools. Our production systems now invoke the Tree-sitter
code directly, rather than being mediated by \semantic{}. The lowered
barrier to entry afforded by the use of the domain-specific query language
allowed the Elixir programming language community to contribute and maintain
their own rules for code navigation, and we anticipate future external
contributions from language communities.

% \section{Future Work}
% \label{futurework}
% \label{sec:future}
\paragraph{Future Plans}
\semantic{} continues development at GitHub, where we are using it to explore
future product features that require high-level program analysis and abstract
interpretation. We have defined compatibility interfaces to bridge
the output of Tree-sitter DSL operations to perform analyses that would be
inconvenient to express in the DSL or without the Haskell ecosystem. We are
now experimenting with abstract interpretation to track exceptions.

Further work remains to generalize the hardcoded \texttt{Err} functor (see
Section \ref{code:codegen}) wrapping child nodes (to handle the fact that Tree-sitter
parsers' error-tolerance means that some subterms may be syntactically invalid)
to a type parameter, rendering them types of kind
$(\star \rightarrow \star) \rightarrow \star $. This approach, thanks to the
flexibility associated with customizing the shape of contained data, can recover
the convenience and flexibility of the unityped version without compromising
type safety. For example, though generated syntax nodes have no fields
representing comments, parameterizing these nodes with a functor containing
comment text will allow us to associate them with comments. Using a three-valued
functor (usually known as \texttt{These}) allows us to recover diffing
capabilities by associating nodes with additions, subtractions, or unchanged
sections in a given diff.

\paragraph{Conclusion}
Even though \semantic{} is no longer part of the code navigation pipeline, we
consider it a successful application of FP techniques, which allowed us to
iterate quickly on a solution, draw from well-researched avenues of thought,
and express complex thoughts concisely, all the while retooling our approach in
the face of evolving business requirements and use cases. We are pleased both
with the utility of \semantic{} as a tool at GitHub’s disposal and as a source
of open-source software and community interest.

\section*{Acknowledgements}
The authors would like to thank all the contributors to \semantic{} over its
lifetime, including Ayman Nadeem, Timothy Clem, Rick Winfrey, Josh Vera, Max
Brunsfeld, Doug Creager, and Charlie Somerville. Many thanks also to the anonymous
ICFP 2022 reviewers for their helpful feedback.

\bibliography{../references}
\clearpage

\end{document}